\documentclass[aps,a4paper,twocolumn,amsmath,amssymb,revsymb,superscriptaddress]{quantumarticle}
\pdfoutput=1
\usepackage{graphicx}
\usepackage[numbers,sort&compress]{natbib}
\usepackage{dcolumn}
\usepackage{bm}
\usepackage{braket}
\usepackage[colorlinks=true, urlcolor=blue, citecolor=blue,
linkcolor=red]{hyperref}
\usepackage{txfonts}
\usepackage{mathtools}

\def\mychi{\raisebox{0.35ex}{$\chi$}}

\def\braket#1{\mathinner{\langle{#1}\rangle}}

\newcommand\symbolwithin[2]{%
  {\mathmakebox[\widthof{\ensuremath{{}#2{}}}][c]{{#1}}}}

\begin{document}

\title{Observing quantum synchronization blockade in circuit quantum electrodynamics}

\author{Simon~E.~Nigg}\email[]{simon.nigg@unibas.ch, usnigg@gmail.com}

\affiliation{Department of Physics, University of Basel,
  Klingelbergstrasse 82, 4056 Basel, Switzerland}
\date{\today}

\begin{abstract}
 High quality factors, strong nonlinearities, and extensive design flexibility make superconducting circuits an ideal
  platform to investigate synchronization phenomena deep in the
  quantum regime. Recently~\cite{Loerch-2017}, it was predicted that energy quantization
  and conservation can block the synchronization of two identical, weakly coupled
  nonlinear self-oscillators. Here we propose a
  Josephson junction circuit realization of such a system along with a
  simple homodyne measurement scheme to observe this effect. We
  also show that at finite detuning, where phase synchronization takes place,
  the two oscillators are entangled in the steady state as witnessed by the positivity of the
  logarithmic negativity. 
\end{abstract}

\maketitle

\section{Introduction}
Synchronization of coupled self-sustained oscillating systems is a
ubiquitous phenomenon in nature
and appears in fields as diverse as biology~\cite{Strogatz-1993, Glass-2001,Chen2017}, economics~\cite{Krause-2015},
sociology~\cite{Bloch-2013} and physics, where it was first
scientifically described~\cite{Huygens-1673}. In the latter field an
interesting question is what happens with synchronization in the
quantum regime, i.e. when the limit cycle steady states of the oscillators are quantum
states with no classical analog. Previous work on quantum
synchronization has focused mainly on theoretically identifying and characterizing
differences between classical and quantum
synchronization~\cite{Zhirov-2006,Zhirov-2009,Mari-2013,Lee-2013,Lee-2014,Walter-2014,Walter-2015,Ameri-2015,Weiss-2016,Fiderer-2016,Loerch-2016,Loerch-2017}
and on potential applications of the latter~\cite{Makino-2016,Giorgi-2016,Bellomo-2017}. Experimental
observation of quantum
synchronization phenomena is hindered by the stringent requirements of
high quantum coherence and strong nonlinearities, both of which are also
key requirements for quantum computation. 
\begin{figure}[t]
  \includegraphics[width=\columnwidth]{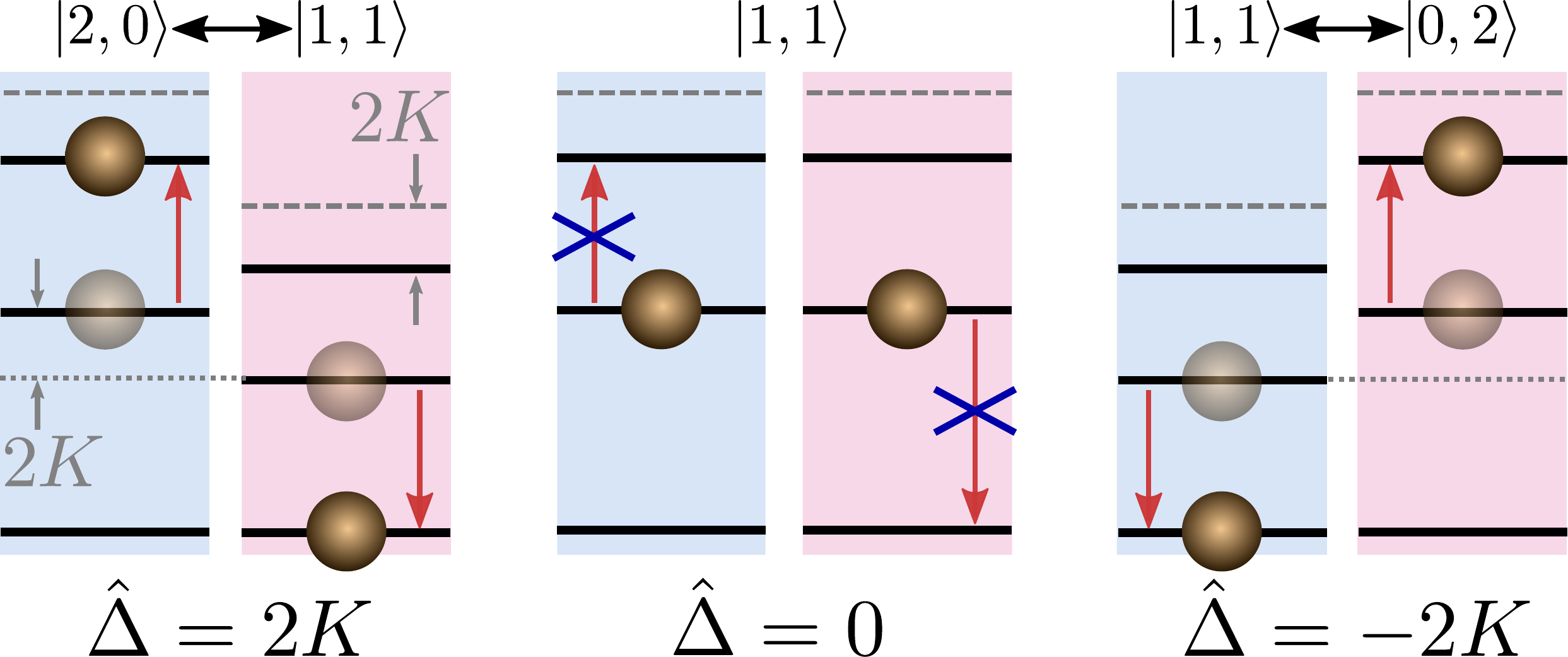}\caption{Energetics
    of quantum synchronization blockade of two weakly
    coupled anharmonic self-oscillators represented as three level
    systems with softening anharmonicity $2K$. A resonance between the
    states $\ket{1,1}$ and $\ket{2,0}$ ($\ket{0,2}$) requires a
    detuning $\hat\Delta=2K$ ($\hat\Delta=-2K$) between the
    oscillators.\label{fig:energetics}}
\end{figure}

Driven in large parts by the quest for a quantum computer, superconducting circuits
realized with one or multiple
Josephson junctions coupled to microwave resonators have become a versatile platform to study
light-matter interaction at the single photon level. 
The design flexibility of superconducting circuits
has enabled the realization of a wide range of Hamiltonians~\cite{Kirchmair-2013a,Devoret-2013,Anderson-2016} and
quantum reservoirs~\cite{Shankar-2013,Kimchi-Schwartz-2016,Liu-2016,Albert-2016} with great precision. This in turn has made
possible the observation of textbook nonlinear quantum optics
effects taking place in hitherto
inaccessible regimes~\cite{Kirchmair-2013a,Toyli-2016}.

Here we show that superconducting circuits form
an ideal platform for studying synchronization of coupled nonlinear
self-sustained oscillators deep in the quantum regime. In particular, we provide
the blueprint of a circuit for observing the {\em quantum
  synchronization blockade} (QSB) recently predicted
by~\citet{Loerch-2017}: In contrast to the classical case where
phase synchronization is maximal between two weakly interacting self-sustained oscillators of equal
frequencies~\cite{Pikovsky}, phase synchronization between two weakly coupled
nonlinear self-oscillators, individually stabilized to a Fock state, is
suppressed on resonance. Intuition for this effect can be
obtained in the perturbative limit of weak interactions~\cite{Loerch-2017} as
illustrated in Fig.~\ref{fig:energetics}: To lowest order, a weak coupling between two
nonlinear self-oscillators stabilized to the Fock state $\ket{1}$ can
only lead to energy exchange when a finite detuning compensates for
the anharmonicity. At zero detuning, energy conservation forbids the
exchange of energy to leading order and synchronization is blocked.

At the heart synchronization is a form of correlation. In the quantum
regime, the relation between synchronization and entanglement is
of
particular
interest~\cite{Zhirov-2009,Manzano-2013,Lee-2014,Witthaut-2017}. This relation
can also be used to define quantum synchronization: If the
correlations present in the synchronized state are non-classical,
i.e. if the two oscillators are in an entangled state, then
synchronization is of quantum origin. Here we show that when
synchronization occurs in our circuit, the steady state of the two
oscillators is
indeed entangled.
  
The focus here is on self-sustained nonlinear quantum oscillators,
which are incoherently driven into non-classical steady states by a
combination of nonlinear damping and amplification (anti-damping). \citet{Rips-2012}
proposed a scheme to stabilize nonlinear nanomechanical oscillators into
non-classical steady states by taking advantage of the quantized
radiation pressure force between microwave photons and a
nanomechanical oscillator.
\begin{figure}[t]
  \includegraphics[width=\columnwidth]{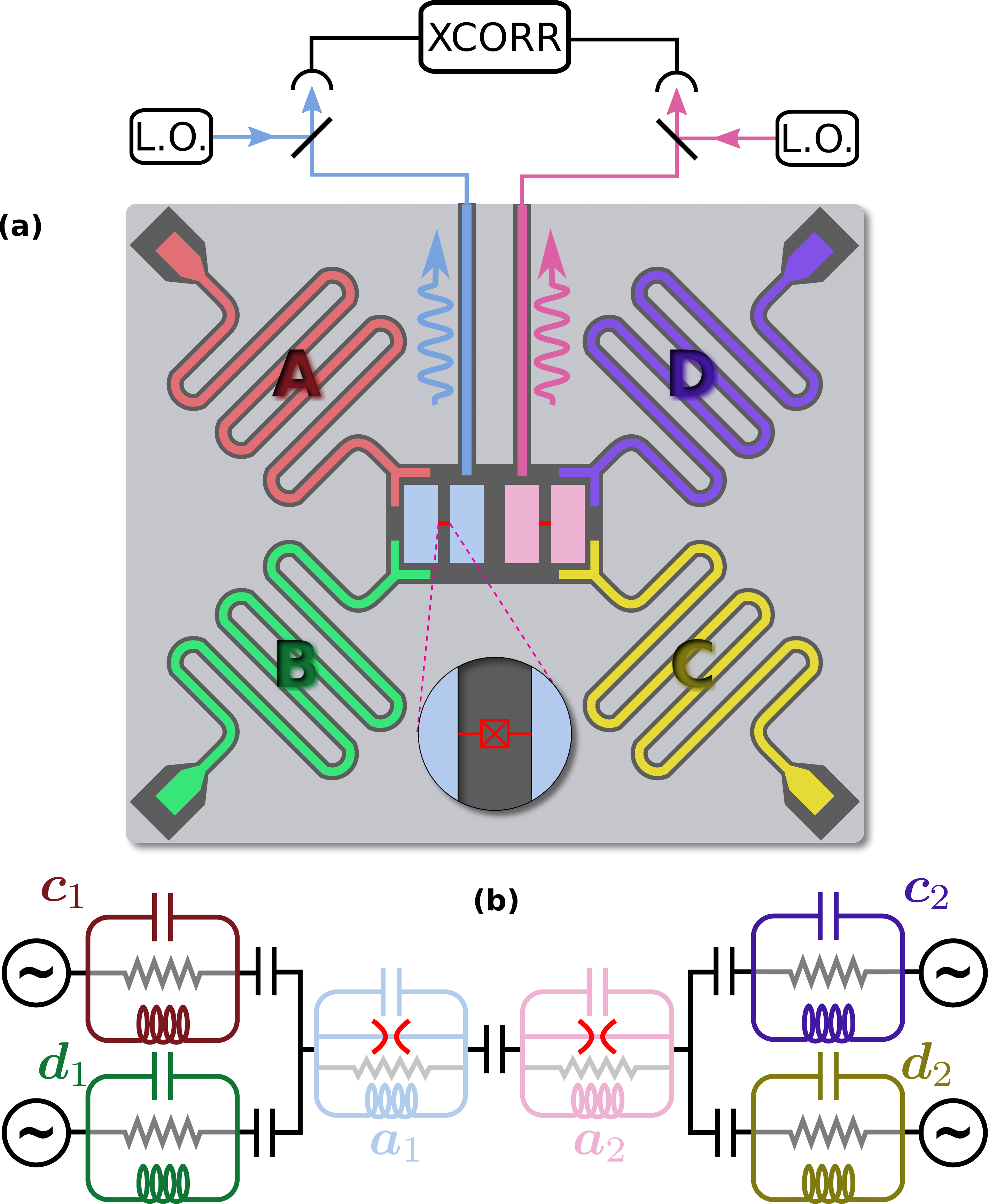}\caption{{\bf (a)} Circuit layout of a coplanar waveguide (CPW)
    realization of the system. The upper part shows the
    measurement setup used to detect synchronization and its blockade
    in the homodyne cross-correlations. {\bf (b)} Minimal lumped element
    circuit model (omitting the readout lines). Frequency multiplexing allows driving of all six modes via
    the four microwave ports shown.\label{fig:setup}}
\end{figure}
Crucially, QSB can be observed only when the energy scale of the nonlinearity $K$
surpasses the damping rate $\kappa$ of the oscillator~\cite{Loerch-2017}. This single
photon Kerr regime, where
$\kappa\ll K$, has not yet
been reached with nanomechanical oscillators currently precluding the
observation of QSB in these
systems. In contrast, owing to tremendous experimental
progress~\cite{Devoret-2013}, superconducting circuits have recently entered this
regime~\cite{Kirchmair-2013a} with ratios as large as $K/\kappa\simeq 10^3$. Thus motivated, we adapt the proposal of \citet{Rips-2012} to
circuit quantum electrodynamics (cQED) replacing the nanomechanical
oscillator with a transmon qubit. The major difficulty in doing so is
that the natural (capacitive or inductive) interaction between
superconducting oscillators is not of the radiation pressure form,
which complicates engineering both nonlinear damping and amplification
simultaneously. One of our key results is that we show
that this can still be approximately achieved in the dispersive regime of cQED in a suitably
displaced and rotated frame thus opening up the possibility to observe
non-classical behavior of quantum synchronization with
state-of-the-art superconducting circuits.\\

\section{Circuit and model}
A coplanar waveguide (CPW) realization of the superconducting circuit we consider is depicted in
Fig.~\ref{fig:setup}~(a) and consists of six oscillator circuits: Two
capacitively coupled Josephson nonlinear oscillators at the center, each capacitively coupled to two linear
microwave resonators. A minimal lumped element circuit model for the relevant
modes (one mode per oscillator) is
shown in Fig.~\ref{fig:setup}~(b). The bare resonance frequencies of all six oscillators are
detuned from each other by many times the coupling strengths. Hence,
in this dispersive regime, where energy exchange is suppressed to first
order, the normal modes of the linearized circuit remain close to the
uncoupled modes and each oscillator can be associated with the corresponding
normal mode as indicated by the color scheme of Fig.~\ref{fig:setup}~(a)
and (b). Interactions are generated by
the two Josephson cosine nonlinearities (red spider symbols in
Fig.~\ref{fig:setup}~(b)), which couple the normal
modes
together~\cite{Nigg-2012a,Bourassa-2012b,Solgun-2014,Smith-2016,Malekakhlagh-2016}. Retaining
only the dominant leading-order terms (see~\cite{supp_mat_qsync} for
a full derivation), the unitary part of the dynamics of
this quantum system is governed
by the Hamiltonian $\bm H=\bm H_{\rm disp}+\bm H_{\rm control}$ with
\begin{widetext}
\begin{align}
& \bm H_{\rm disp}=\sum_{j=1}^2\left( \Delta^{a}_j\bm a_j^{\dagger}\bm
  a_j^{}+\Delta^{c}_j\bm c_j^{\dagger}\bm c_j^{}+\Delta^{d}_j\bm
  d_{j}^{\dagger}\bm d_j^{}-K_j\bm a_j^{\dagger}\bm a_j^{\dagger}\bm
  a_j^{}\bm a_j^{}-\mychi_j^{ac}\bm a_j^{\dagger}\bm a_j^{}\bm c_j^{\dagger}\bm
  c_j^{}-\mychi^{ad}_j\bm a_j^{\dagger}\bm a_j^{}\bm d_j^{\dagger}\bm
  d_j^{}\right) + J\bm a_1^{\dagger}\bm a_1^{}\bm a_2^{\dagger}\bm
  a_2^{},\label{eq:2}\\
                  &\bm H_{\rm
                  control}=\sum_{j=1}^2\left\{ \varepsilon_j^{a}\left( \bm
                            a_j^{}+\bm a_j^{\dagger} \right) +
                            \varepsilon_j^{c}\left( \bm c_j^{}+\bm
                            c_j^{\dagger} \right)
                            +\varepsilon_j^{d}\left( \bm d_j^{}+\bm d_j^{\dagger} \right)\right\}.\label{eq:3}
\end{align}
\end{widetext}
Here $\bm a_i$ is the bosonic annihilation operator of the normal mode
associated with the nonlinear oscillator $i\in\{1,2\}$, which is coupled to two linear resonators
with associated normal mode operators $\bm c_i$ and $\bm
d_i$. This Hamiltonian is written in a frame rotating with the drives
such that $\Delta^{a}_j,\Delta^{c}_j$ and $\Delta^{d}_j$ denote the detunings
between the corresponding resonators and drives. The rotating wave
approximation (RWA) has been applied and the dominant interactions between the modes present in Eq.~(\ref{eq:2}), which stem from the $\bm\varphi^4$
term of the Josephson potentials, are of self-Kerr ($K_j$)
and cross-Kerr ($J$, $\mychi_j^{ac}$ and $\mychi_j^{ad}$) form.

The other crucial ingredient to achieve the desired limit cycle steady state is dissipation due to photon
losses, which, in typical quantum optics fashion, is captured via the zero temperature Lindblad
master equation
\begin{align}\label{eq:6}
  \bm{\dot\rho}&=-i[\bm
  H,\bm\rho]\\
  &+\sum_{i=1,2}\left( \kappa_i^{a}\mathcal{D}[\bm
a_i]+\kappa_i^{c}\mathcal{D}[\bm c_i] +\kappa_i^{d}\mathcal{D}[\bm
  d_i] \right)\bm\rho,\nonumber
\end{align}
with dissipator $\mathcal{D}[\bm O]\bm\rho=\bm O\bm\rho\bm
  O^{\dagger}-\frac{1}{2}\{\bm O^{\dagger}\bm O,\bm\rho\}$, where
$\{\bm A,\bm B\}=\bm A\bm B+\bm B\bm A$. Here $\kappa_i^{a}$
describes photon losses in the nonlinear oscillator $i$ and in
Section~\ref{sec:homodyne-detect-quan} we will use this channel for
measurements. The rates $\kappa_i^{c}$ and $\kappa_i^{d}$ account for
photon losses in the linear resonators and we shall assume
$\kappa_i^a\ll\kappa_i^c,\kappa_i^d$, as motivated further below.
\section{Fock state stabilization ($J=0$)}
To enter the regime of quantum synchronization
blockade the nonlinear self-oscillators need to be stabilized to a Fock
state~\cite{Loerch-2017}. Fock state stabilization in linear superconducting oscillators
has been achieved in~\cite{Holland-2015} by means of an autonomous feedback
mechanism mediated by a strong dispersive interaction with a
qubit. Here we show that the system modeled by Eqs.~(\ref{eq:2}--\ref{eq:6}), with $J=0$ can be used to
stabilize a Fock state in the nonlinear oscillators. To do so we
adapt the proposal of \citet{Rips-2012} for stabilizing a single
phonon Fock state of a nonlinear nano-mechanical oscillator to the stabilization
of a photonic Fock state
of a nonlinear superconducting
oscillator. Details of our derivation are
provided in~\cite{supp_mat_qsync}. In the following we sketch the main
steps focusing on the differences with~\cite{Rips-2012}.
\begin{figure}[ht]
  \includegraphics[width=\columnwidth]{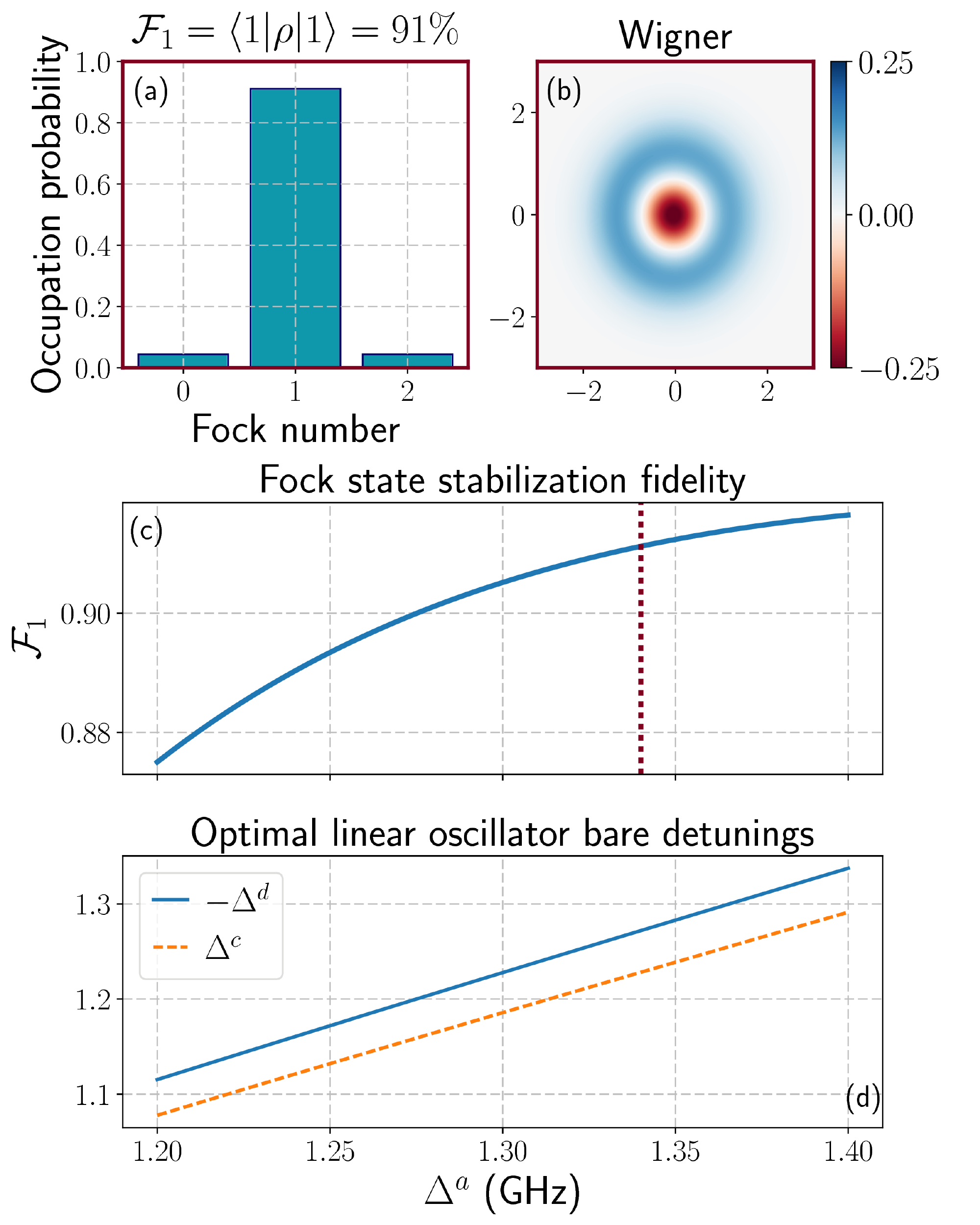}\caption{Fock
    state stabilization. {\bf (a)} Photon number distribution of the
    nonlinear oscillator in the
    steady state. {\bf (b)} Wigner function of the steady
    state. {\bf (c)} Fidelity of the stabilized state to the Fock
    state $\ket{1}$ as a function of the nonlinear oscillator
    detuning. {\bf (d)} Detunings of the linear oscillators for
    optimal stabilization. These results are obtained by numerically integrating the
    master equation~(\ref{eq:6}) with $J=0$ and parameters
    $\kappa^{a}=100\,{\rm kHz}$, $\kappa^{c}=\kappa^{d}=10\,{\rm
    MHz}$, $K=30\,{\rm MHz}$, $\mychi^{ac}=\mychi^{ad}=8\,{\rm MHz}$,
  $\varepsilon^{a}=500\,{\rm MHz}$ and
  $\varepsilon^{c}=\varepsilon^{d}=2\,{\rm GHz}$.\label{fig:fock_stab}}
\end{figure}

Since $J=0$, we can concentrate on the subsystem consisting of oscillator modes
$\bm a_1$, $\bm c_1$ and $\bm d_1$ without loss of generality and for
compactness we drop the subscript $1$ in what follows. The central new
idea here is
to use the drive terms in $\bm H_{\rm control}$ to coherently displace
the modes, i.e. $\bm a\rightarrow\bm a +\alpha$, $\bm c\rightarrow\bm
c+\gamma$ and $\bm d\rightarrow\bm d+\delta$, such as to generate, via the cross-Kerr terms, a Rabi-type coupling between the linear
and nonlinear oscillators:
\begin{align}
\mychi^{ac}\bm a^{\dagger}\bm a\bm c^{\dagger}\bm c&+\mychi^{ad}\bm
  a^{\dagger}\bm a\bm d^{\dagger}\bm d\label{eq:4}\\
  &\symbolwithin{\downarrow}{+} \nonumber\\
  \mychi^{ac}\left(\alpha^*\bm a+\alpha\bm
  a^{\dagger}\right)&\left(\gamma^*\bm c+\gamma\bm c^{\dagger}\right)+
                                                   \nonumber\\
 \mychi^{ad}\left(
  \alpha^*\bm a+\alpha\bm a^{\dagger} \right)&\left( \delta^*\bm d+\delta\bm d^{\dagger} \right)+\dots\label{eq:5}
\end{align}
The displacement amplitudes are given by
$\alpha=-\varepsilon^{a}/(\Delta^{a}-i\kappa^{a}/2)$,
$\gamma=-\varepsilon^{c}/(\Delta^{c}-i\kappa^{c}/2)$ and
$\delta=-\varepsilon^{d}/(\Delta^{d}-i\kappa^{d}/2)$. These
amplitudes are chosen such as to cancel the drive terms upon
displacing the quadratic and dissipative terms. The displacement generates additional terms indicated by the ellipses in
Eq.~(\ref{eq:5}). The effect of most of these terms can be neglected
in RWA, but some terms are non-rotating and
lead to renormalizations of the oscillator frequencies:
$\Delta^{a}\rightarrow\tilde\Delta^{a}=\Delta^{a}-\mychi^{ac}|\gamma|^2-\mychi^{ad}|\delta|^2$
and $\Delta^{c}\rightarrow\tilde\Delta^{c}=\Delta^{c}-\mychi^{ac}|\alpha|^2$ as well
as
$\Delta^{d}\rightarrow\tilde\Delta^{d}=\Delta^{d}-\mychi^{ad}|\alpha|^2$. In
addition, the
displacement transformation of the Kerr term of the $\bm a$
mode generates two more non-rotating terms: The first one leads to an additional
frequency renormalization:
$\tilde\Delta^{a}\rightarrow\hat\Delta^{a}=\tilde\Delta^{a}-4K|\alpha|^2$ and
the second one is the squeezing term $K(\alpha^{*2}\bm a^2+\alpha^2\bm
a^{\dagger 2})$. The latter can
potentially adversely affect Fock state stabilization. However, we find
that its effect on the steady state is negligible if the displacement
amplitude $\alpha$ of the nonlinear oscillator remains small compared
with the displacement amplitudes $\gamma$ and $\delta$ of the linear
oscillators such that $K/\mychi^{ac}\ll|\gamma/\alpha|$ as well as
$K/\mychi^{ad}\ll|\delta/\alpha|$.
\begin{figure*}[ht]
  \includegraphics[width=\textwidth]{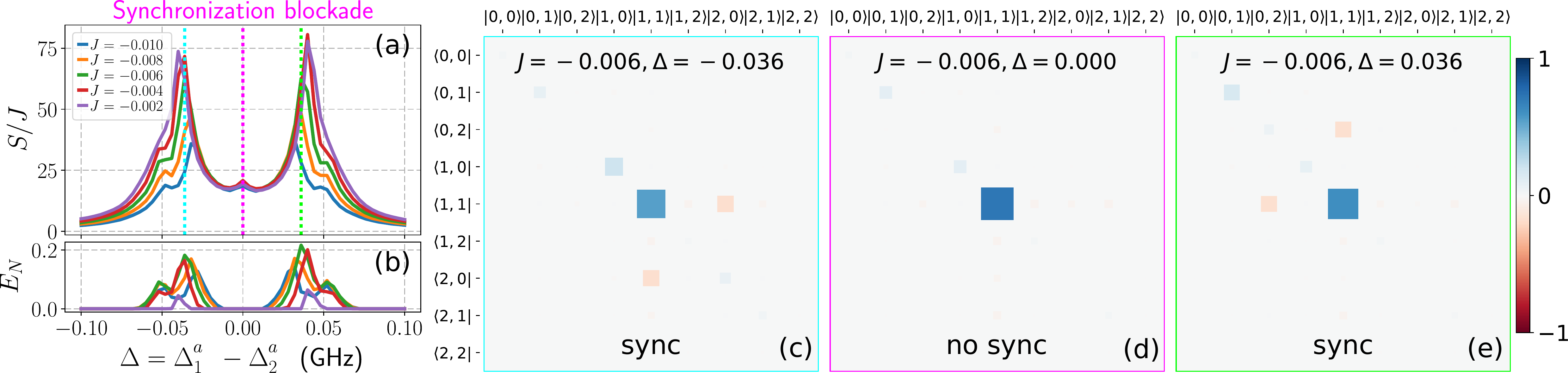}\caption{{\bf
    (a)} Normalized quantum synchronization measure $S/J$ as a function of
  the bare detuning $\Delta=\Delta_1^{a}-\Delta_2^{a}$ between the two nonlinear
  self-oscillators for different values of the bare inter-oscillator
  dispersive coupling strength $J$. $\Delta_2^{a}$ is varied while $\Delta_1^{a}$
  is kept fixed. The steady state is obtained from a quantum
  trajectory simulation by averaging the long time (i.e. $t\gg
  1/\gamma_{\pm}$) temporal averages over $500$ trajectories. {\bf
  (b)} Logarithmic negativity $E_N={\rm log_2}||\bm\rho_{\rm
  ss}^{\rm PT} ||_1$, showing that when
the oscillators synchronize, entanglement is present,
i.e. $E_N>0$. {\bf (c)} to {\bf (e)} Hinton diagrams at three different
  values of $\Delta$ showing off-resonant synchronization (c) and (e) and
  synchronization blockade on resonance (d). The area of the squares
  is proportional to the amplitude of the complex number $\braket{k,l|\bm\rho_{\rm
      ss}|m,n}$ in Fock space and the color corresponds to its real part. Correlations between the states $\ket{1,1}$ and
  $\ket{2,0}$ respectively $\ket{0,2}$ are clearly visible off resonance ((c) and
  (e)) but are absent on resonance (d). Animations over the full range
  of detunings are provided at this url~\cite{supp_mat_qsync}. Parameter values are the same as in Fig.~\ref{fig:fock_stab}.\label{fig:sync}}
\end{figure*}

Together with appropriately detuned drives, the interaction of
Eq.~(\ref{eq:5}), allows a state with a narrow excitation number distribution
centered around $n_0\in\mathbb{N}$ to be stabilized. Specifically this requires that
$\hat\Delta^{a}=\tilde\Delta^{c}+\Delta_{\downarrow}$ as well as
$\hat\Delta^{a}=-\tilde\Delta^{d}+\Delta_{\uparrow}$, where
$\Delta_{\downarrow}=2Kn_0\ll|\hat\Delta^{a}+\tilde\Delta^c|$ and
$\Delta_{\uparrow}=2K(n_0-1)\ll|\hat\Delta^{a}-\tilde\Delta^d|$. Under these
conditions, the red and blue sideband terms
$\mychi^{ac}\alpha^*\gamma\bm a^{\dagger}\bm c
+{\rm H.c.}$, and $\mychi^{ad}\alpha^*\delta^*\bm a\bm d+{\rm H.c.}$,
are made simultaneously resonant, while the counter-rotating terms
$\mychi_{ac}\alpha\gamma \bm a\bm c+{\rm H.c.}$ and
$\mychi_{ad}\alpha\delta^*\bm a^{\dagger}\bm d+{\rm H.c.}$ can be
neglected in RWA. When
$|\alpha\gamma\mychi^{ac}|,|\alpha\delta\mychi^{ad}|\ll\kappa^{c},\kappa^{d}$,
the modes of the linear oscillators can further be adiabatically eliminated
yielding an effective master
equation with a Lorentzian spectrum for a Kerr oscillator with nonlinear damping and
amplification~\cite{supp_mat_qsync,Rips-2012,Loerch-2017}:
\begin{align}\label{eq:7}
  \bm{\dot\rho}&=-i[\bm H_K,\bm\rho]+\kappa^{a}\mathcal{D}\left[ \bm
                 a\right]\bm\rho\\
  &+\gamma_{\uparrow}\sum_mL_{m}\mathcal{D}\left[ \sqrt{m}\ket{m}\bra{m-1}
    \right]\bm\rho\nonumber\\
&+\gamma_{\downarrow}\sum_mL_{m}\mathcal{D}\left[ \sqrt{m+1}\ket{m}\bra{m+1}
  \right]\bm\rho.    \nonumber
\end{align}
Here $\gamma_{\uparrow}=4|\alpha\delta\mychi^{ad}|^2/\kappa$, $\gamma_{\downarrow}=4|\alpha\gamma\mychi^{ac}|^2/\kappa$, $L_m=\sigma^2/[(n_0-m)^2+\sigma^2]$, $\sigma=\kappa/(4K)$ and $\bm H_K=\hat\Delta^{a}\bm a^{\dagger}\bm a-K\bm
a^{\dagger}\bm a^{\dagger}\bm a\bm a$. For compactness we
consider the case where
$\kappa^{c}=\kappa^{d}=\kappa$. When $\kappa\ll 2K$, the
dominant transitions are $n_0-1\rightarrow n_0$ at rate
$\gamma_{\uparrow}L_{n_0}$ and $n_0+1\rightarrow n_0$ at rate
$\gamma_{\downarrow}L_{n_0}$. The system is thus stabilized in the Fock state
$\ket{n_0}$, which is an eigenstate of $\bm
H_K$.

Fig.~\ref{fig:fock_stab} shows the results of a
numerical simulation with the full model with $J=0$
(Eqs.~(\ref{eq:2}--\ref{eq:6})), which
includes all the counter-rotating terms. Because of the
frequency renormalization, we cannot directly obtain the values of the
bare detunings. Instead, for each value of $\Delta^{a}$ we determine
the corresponding values of $\Delta^{c}$ and $\Delta^{d}$ by
maximizing the fidelities to the target Fock state (Fig.~\ref{fig:fock_stab}~(d)). The achievable
fidelities (Fig.~\ref{fig:fock_stab}~(c)) range roughly between $88\%$ and $92\%$ and increase with
increasing detuning of the drive oscillators consistent with the
RWA. Thus we have established that the Fock state stabilizing dynamics of
Eq.~(\ref{eq:7}) can be engineered in our circuit. We next turn to the
case of two coupled systems, i.e. $J\not=0$.
\section{Quantum synchronization blockade}
Classically, two weakly coupled self-oscillators display phase
synchronization, i.e. the relative phase between the two oscillators
is narrowly distributed around a fixed value in the steady state~\cite{Pikovsky}. This
synchronization is strongest on resonance, i.e. when the two isolated
systems oscillate at the same frequency. In contrast,
\citet{Loerch-2017} predict that if two identical Fock state stabilized Kerr oscillators
are weakly coupled with each other by a linear term of the form $J(\bm
a_1^{\dagger}\bm a_2^{}+{\rm H.c.})$, phase synchronization is suppressed
at zero detuning in stark contrast to the classical situation.

In our system (Eqs.~(\ref{eq:2}) and~(\ref{eq:3})), a linear coupling is obtained naturally in the
displaced frame when $J\not=0$, since $J\bm a_1^{\dagger}\bm a_1^{}\bm a_2^{\dagger}\bm
a_2\rightarrow J\left( \alpha_1\alpha_2^*\bm a_1^{\dagger}\bm
a_2^{}+\alpha_1^*\alpha_2\bm a_2^{\dagger}\bm a_1^{} \right)+\dots$, where the
ellipses denote additional terms. Among the latter the only relevant
ones under RWA are $J|\alpha_1|^2\bm a_2^{\dagger}\bm a_2^{}$ and
$J|\alpha_2|^2\bm a_1^{\dagger}\bm a_1^{}$, which lead to an
additional frequency renormalization.

To quantify phase
synchronization we follow~\cite{Hush-2015, Loerch-2017} and use the measure
\begin{align}\label{eq:8}
  S=2\pi\,\underset{\phi}{\rm max}\left[ P(\phi) \right]-1,
\end{align}
with $P(\phi)=\int_0^{2\pi}d\phi_1d\phi_2\delta(\phi_1-\phi_2-\phi)p(\phi_1,\phi_2)$
and $p(\phi_1,\phi_2)=\braket{\phi_1,\phi_2|\bm\rho_{\rm
    ss}|\phi_1,\phi_2}$ and $\ket{\phi_i}=(2\pi)^{-1}\sum_{n=0}^{\infty}e^{in\phi_i}\ket{n}$.
Here $\bm\rho_{\rm ss}$ denotes the steady state reduced density matrix of the
two nonlinear oscillators. The quantity $S\geq 0$ essentially measures how
\begin{figure*}[t]
  \includegraphics[width=\textwidth]{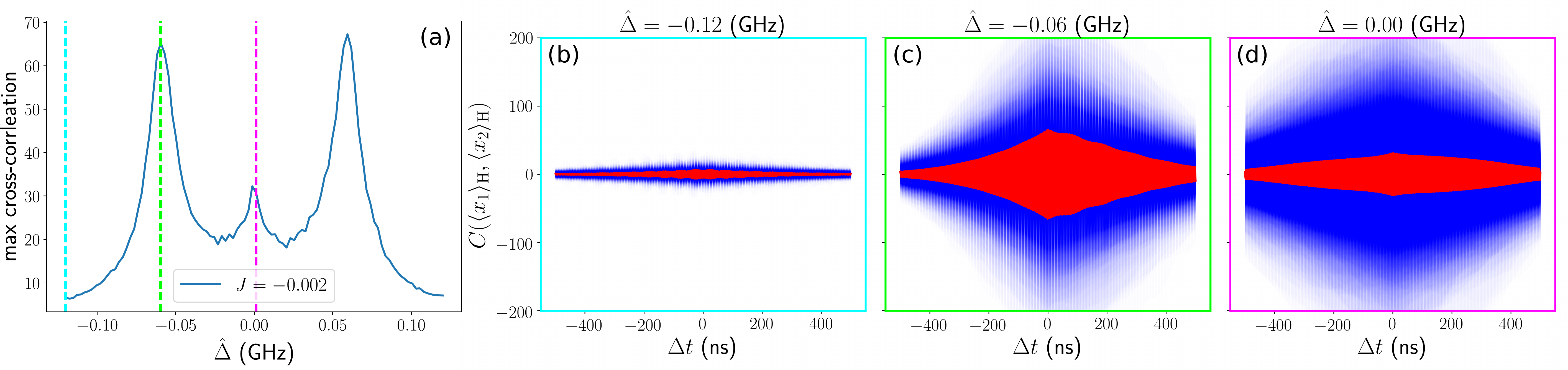}\caption{{\bf
      (a)} Maximum of the averaged cross-correlated homodyne
    signals as a function of
    the renormalized detuning $\hat\Delta$ between the nonlinear oscillators. Clearly the
    cross-correlation mirrors the synchronization measure of
    Fig.~\ref{fig:sync}~(a). {\bf (b)} to {\bf (d)} Cross-correlation
    functions of the homodyne signals corresponding to the three
    different values of $\hat\Delta$ marked by dashed vertical lines in
    (a). The individual trajectories are shown in blue and the average
    over $1000$ trajectories is shown in red. Parameter values
    are the same as in Figs.~\ref{fig:fock_stab} and~\ref{fig:sync}.\label{fig:xcorr}}
\end{figure*}
uniform the relative phase distribution of the two oscillators is. In the absence of synchronization $S=0$ and
the larger $S$ is, the stronger the synchronization. Fig.~\ref{fig:sync} confirms that quantum
synchronization blockade does indeed occur in our
system as reflected by a suppression of $S$ on resonance
(Fig.~\ref{fig:sync}~(a)). At finite detuning, when the resonance
condition $\hat\Delta\equiv\hat\Delta_1^{a}-\hat\Delta_2^{a}=\pm 2K$ is satisfied,
the phases of the oscillators synchronize as indicated by the
presence of peaks in $S$ (Fig.~\ref{fig:sync}~(a)). Notice that the spacing between
the two major peaks is reduced compared with the ideal
separation of $4K$. This is due to the frequency
renormalization $\Delta\rightarrow\hat\Delta$ discussed above and to
the coupling, which split the resonances by $\pm J$. As the coupling strength $J$
increases, off-resonant synchronization is reduced and a small bump appears on
resonance. Also,
because the fidelity of Fock state stabilization (for $J=0$) depends
on the detuning (Fig.~\ref{fig:fock_stab}~(c)), the synchronization signal is slightly
stronger for positive than for negative detunings.

Interestingly, when the two oscillators synchronize, their state is
entangled as witnessed by the positivity of the logarithmic
negativity~\cite{Plenio-2005} $E_N={\rm log_2}||
    \bm\rho_{\rm ss}^{\rm PT} ||_1  $ (Fig.~\ref{fig:sync}~(b)). Here
    $\bm\rho_{\rm ss}^{\rm PT}$ denotes the steady state density matrix partially transposed with respect to
one of the two oscillators and $||\bm \rho||_1={\rm
  tr}\sqrt{\bm\rho\bm\rho^{\dagger}}$ is the trace norm. To gain further
insight into the nature of the blocked and the synchronized states we
plot Hinton diagrams of $\bm\rho_{\rm ss}$ for $J=-6\,{\rm MHz}$ for
the resonant case $\Delta=0$ (Fig.~\ref{fig:sync}~(d)) and for the two
bare detunings $\Delta\simeq\pm 0.036\,{\rm GHz}$, which correspond to a
synchronization resonance (Fig.~\ref{fig:sync}~(c) and (e)). When $\Delta=0$, the two oscillators are
essentially in the tensor product state $\ket{1,1}$. For
$\Delta=-0.036\,{\rm GHz}$ ($\Delta=0.036\,{\rm GHz}$) this state hybridizes with the state
$\ket{0,2}$ ($\ket{2,0}$) resulting ideally in the entangled doublet
$c_{11}\ket{1,1}\pm c_{02}\ket{0,2}$ ($c_{11}\ket{1,1}\pm
c_{20}\ket{2,0}$). The amplitudes $c_{11}$, $c_{20}$ and $c_{02}$ are
determined by the competition between the localizing Fock state stabilization and
the delocalizing inter-oscillator coupling. Since here the coupling is weak
the former dominates leading to a relatively small by-mixing of
the states $\ket{2,0}$ and $\ket{0,2}$. Animations illustrating the synchronization blockade and the
splitting of the resonances as $\Delta$ is varied can be found at this url~\cite{supp_mat_qsync}.

\section{Homodyne detection of quantum synchronization and its blockade\label{sec:homodyne-detect-quan}}
While the synchronization measure $S$ is convenient for theoretical
characterization, it can be challenging to measure since it requires
full state tomography of the two nonlinear oscillators. Here we show
that quantum synchronization and its blockade can be detected more
simply by correlating the classical signals obtained from separate
homodyne measurements of the phases of the nonlinear oscillators.

We choose the phases of the local oscillators such as to measure the
$\bm X_i=\bm a_i+\bm a_i^{\dagger}$ quadratures. The
homodyne signals from the two oscillators are then given by~\cite{Wiseman_Milburn-2009}
\begin{align}
  J_i(t)= \kappa_i^{a}\braket{\bm X_i}_{\rm H}+\sqrt{\kappa_i^{a}}\xi_i(t),
\end{align}
where $\xi_i(t)$ is a zero mean Gaussian white noise random process
with unit variance and the
subscript ${\rm H}$ indicates that the expectation value is
conditioned on the particular quantum trajectory $\bm\rho_{\rm H}(t)$. The latter is a
solution of the stochastic master equation
\begin{align}
\bm{\dot\rho}_{\rm H}&=\mathcal{\bm L}_K\bm\rho_{\rm H}+\sum_{i=1,2}\left(
                       J_i-\kappa_i^a\braket{\bm X_i}_{\rm H} \right)\mathcal{H}[\bm
  a_i]\bm\rho_{\rm H}.
\end{align}
Here $\mathcal{\bm L}_K$ is the Liouvillian superopertor generating the right-hand
side of Eq.~(\ref{eq:7}), $\kappa_i^a$ plays the role of the coupling to measurement device and
$\mathcal{H}[\bm O]\bm\rho=\bm O\bm\rho+\bm\rho\bm O^{\dagger}-{\rm
  tr}[\bm O\bm\rho+\bm \rho\bm O^{\dagger}]\bm\rho$, describes the
back-action of the homodyne measurement. The phase correlation
between the two oscillators is reflected in the averaged cross-correlation
of the measurement signals. Specifically we look at the
maximum of the averaged cross-correlation given by
\begin{align}
\mathbb{E}\left[ C_{\tau}(J_1, J_2) \right] =
  (\kappa_1^{a}\kappa_2^{a})\mathbb{E}\left[
  C_{\tau}(\braket{X_1}_{\rm H}, \braket{X_2}_{\rm H}) \right],
\end{align}
where $C_{\tau}(x,y)=\int_0^{T}x(t)y(t-\tau)dt$ is the
cross-correlation function, $T\gg \gamma_{\uparrow}^{-1},\gamma_{\downarrow}^{-1}$ is the
measurement time, and $\mathbb{E}[\cdot]$ denotes the average value over
trajectories. Fig.~\ref{fig:xcorr} shows the results of a stochastic
master equation simulation using the effective model~(\ref{eq:7}) for
each oscillator with the same parameter values as used in
Figs.~\ref{fig:fock_stab} and~\ref{fig:sync}. This confirms that phase synchronization and
its blockade as quantified by $S$ (Eq.~(\ref{eq:8})), are indeed reflected in the cross-correlations of the
homodyne signals. It is worthwhile to emphasize the
simplicity of this detection scheme and we hope that its adoption will
accelerate progress on the experimental front.

\section{Conclusions}
Bath engineering with state-of-the-art superconducting circuits
together with measurement methods typically used for
quantum information processing offer a versatile approach to
investigate quantum synchronization phenomena. We proposed and
analyzed a circuit to test the recently predicted phenomenon of
quantum synchronization blockade~\cite{Loerch-2017}. Its realization
should be within reach of current technology. We also showed that
off-resonant quantum synchronization in this system is accompanied by entanglement. Furthermore, we
showed that quantum synchronization and its blockade can be detected in
cross-correlations of homodyne signals, greatly simplifying their
experimental identification. It is expected that the
results presented here will also be relevant in the near future when
it becomes possible to build larger networks of
high-Q nonlinear superconducting oscillators. Such systems are
predicted to display
a rich variety of quantum effects~\cite{Benedetti-2016,Loerch-2017}
and have potential technological applications~\cite{Makino-2016,Li-2017,Nigg-2017,Puri-2017}.

{\it Acknowledgments}: This work was supported by the Swiss National
Science Foundation (SNSF) and the NCCR QSIT. We acknowledge
interesting discussions with Niels L{\"o}rch, Andreas Nunnenkamp, Rakesh Tiwari and Christoph Bruder. The numerical computations were peformed in a
parallel computing environment at sciCORE (\url{http://scicore.unibas.ch/}) scientific computing core facility
at University of Basel using the Python libraries NumPy
(\url{http://www.numpy.org/}) and QuTip (\url{http://qutip.org/}). The
graphics were created with
Matplotlib (\url{http://www.matplotlib.org}).

\bibliographystyle{apsrev}
\bibliography{/home/sen/phys/library/bibtex/mybib}

\begin{thebibliography}{47}
\expandafter\ifx\csname natexlab\endcsname\relax\def\natexlab#1{#1}\fi
\expandafter\ifx\csname bibnamefont\endcsname\relax
  \def\bibnamefont#1{#1}\fi
\expandafter\ifx\csname bibfnamefont\endcsname\relax
  \def\bibfnamefont#1{#1}\fi
\expandafter\ifx\csname citenamefont\endcsname\relax
  \def\citenamefont#1{#1}\fi
\expandafter\ifx\csname url\endcsname\relax
  \def\url#1{\texttt{#1}}\fi
\expandafter\ifx\csname urlprefix\endcsname\relax\def\urlprefix{URL }\fi
\providecommand{\bibinfo}[2]{#2}
\providecommand{\eprint}[2][]{\url{#2}}

\bibitem[{\citenamefont{Strogatz and Stewart}(1993)}]{Strogatz-1993}
\bibinfo{author}{\bibfnamefont{S.~H.} \bibnamefont{Strogatz}} \bibnamefont{and}
  \bibinfo{author}{\bibfnamefont{I.}~\bibnamefont{Stewart}},
  \bibinfo{journal}{Scientific American} \textbf{\bibinfo{volume}{269}},
  \bibinfo{pages}{102} (\bibinfo{year}{1993}).

\bibitem[{\citenamefont{Glass}(2001)}]{Glass-2001}
\bibinfo{author}{\bibfnamefont{L.}~\bibnamefont{Glass}},
  \bibinfo{journal}{Nature} \textbf{\bibinfo{volume}{410}},
  \bibinfo{pages}{277} (\bibinfo{year}{2001}), ISSN \bibinfo{issn}{0028-0836},
  \urlprefix\url{http://dx.doi.org/10.1038/35065745}.

\bibitem[{\citenamefont{Chen et~al.}(2017)\citenamefont{Chen, Liu, Shi,
  Chat{\'e}, and Wu}}]{Chen2017}
\bibinfo{author}{\bibfnamefont{C.}~\bibnamefont{Chen}},
  \bibinfo{author}{\bibfnamefont{S.}~\bibnamefont{Liu}},
  \bibinfo{author}{\bibfnamefont{X.-q.} \bibnamefont{Shi}},
  \bibinfo{author}{\bibfnamefont{H.}~\bibnamefont{Chat{\'e}}},
  \bibnamefont{and} \bibinfo{author}{\bibfnamefont{Y.}~\bibnamefont{Wu}},
  \bibinfo{journal}{Nature} \textbf{\bibinfo{volume}{542}},
  \bibinfo{pages}{210} (\bibinfo{year}{2017}), ISSN \bibinfo{issn}{0028-0836},
  \bibinfo{note}{letter},
  \urlprefix\url{http://dx.doi.org/10.1038/nature20817}.

\bibitem[{\citenamefont{Krause et~al.}(2015)\citenamefont{Krause, B\"orries,
  and Bornholdt}}]{Krause-2015}
\bibinfo{author}{\bibfnamefont{S.~M.} \bibnamefont{Krause}},
  \bibinfo{author}{\bibfnamefont{S.}~\bibnamefont{B\"orries}},
  \bibnamefont{and}
  \bibinfo{author}{\bibfnamefont{S.}~\bibnamefont{Bornholdt}},
  \bibinfo{journal}{Phys. Rev. E} \textbf{\bibinfo{volume}{92}},
  \bibinfo{pages}{012815} (\bibinfo{year}{2015}),
  \urlprefix\url{https://link.aps.org/doi/10.1103/PhysRevE.92.012815}.

\bibitem[{\citenamefont{Bloch et~al.}(2013)\citenamefont{Bloch, Herzog, Levine,
  and Schwartz}}]{Bloch-2013}
\bibinfo{author}{\bibfnamefont{G.}~\bibnamefont{Bloch}},
  \bibinfo{author}{\bibfnamefont{E.~D.} \bibnamefont{Herzog}},
  \bibinfo{author}{\bibfnamefont{J.~D.} \bibnamefont{Levine}},
  \bibnamefont{and} \bibinfo{author}{\bibfnamefont{W.~J.}
  \bibnamefont{Schwartz}}, \bibinfo{journal}{Proceedings of the Royal Society
  of London B: Biological Sciences} \textbf{\bibinfo{volume}{280}}
  (\bibinfo{year}{2013}), ISSN \bibinfo{issn}{0962-8452},
  \eprint{http://rspb.royalsocietypublishing.org/content/280/1765/20130035.full.pdf},
  \urlprefix\url{http://rspb.royalsocietypublishing.org/content/280/1765/20130035}.

\bibitem[{\citenamefont{Hugenii}(1673)}]{Huygens-1673}
\bibinfo{author}{\bibfnamefont{C.}~\bibnamefont{Hugenii}},
  \emph{\bibinfo{title}{Horoloquim Oscilatorium}} (\bibinfo{year}{1673}),
  \bibinfo{note}{english translation, {\it The Pendulum Clock} (Iowa State
  University Press, Ames, Iowa, 1986)}.

\bibitem[{\citenamefont{{Zhirov, O. V.} and {Shepelyansky, D.
  L.}}(2006)}]{Zhirov-2006}
\bibinfo{author}{\bibnamefont{{Zhirov, O. V.}}} \bibnamefont{and}
  \bibinfo{author}{\bibnamefont{{Shepelyansky, D. L.}}}, \bibinfo{journal}{Eur.
  Phys. J. D} \textbf{\bibinfo{volume}{38}}, \bibinfo{pages}{375}
  (\bibinfo{year}{2006}),
  \urlprefix\url{https://doi.org/10.1140/epjd/e2006-00011-9}.

\bibitem[{\citenamefont{Zhirov and Shepelyansky}(2009)}]{Zhirov-2009}
\bibinfo{author}{\bibfnamefont{O.~V.} \bibnamefont{Zhirov}} \bibnamefont{and}
  \bibinfo{author}{\bibfnamefont{D.~L.} \bibnamefont{Shepelyansky}},
  \bibinfo{journal}{Phys. Rev. B} \textbf{\bibinfo{volume}{80}},
  \bibinfo{pages}{014519} (\bibinfo{year}{2009}),
  \urlprefix\url{https://link.aps.org/doi/10.1103/PhysRevB.80.014519}.

\bibitem[{\citenamefont{Mari et~al.}(2013)\citenamefont{Mari, Farace, Didier,
  Giovannetti, and Fazio}}]{Mari-2013}
\bibinfo{author}{\bibfnamefont{A.}~\bibnamefont{Mari}},
  \bibinfo{author}{\bibfnamefont{A.}~\bibnamefont{Farace}},
  \bibinfo{author}{\bibfnamefont{N.}~\bibnamefont{Didier}},
  \bibinfo{author}{\bibfnamefont{V.}~\bibnamefont{Giovannetti}},
  \bibnamefont{and} \bibinfo{author}{\bibfnamefont{R.}~\bibnamefont{Fazio}},
  \bibinfo{journal}{Phys. Rev. Lett.} \textbf{\bibinfo{volume}{111}},
  \bibinfo{pages}{103605} (\bibinfo{year}{2013}),
  \urlprefix\url{https://link.aps.org/doi/10.1103/PhysRevLett.111.103605}.

\bibitem[{\citenamefont{Lee and Sadeghpour}(2013)}]{Lee-2013}
\bibinfo{author}{\bibfnamefont{T.~E.} \bibnamefont{Lee}} \bibnamefont{and}
  \bibinfo{author}{\bibfnamefont{H.~R.} \bibnamefont{Sadeghpour}},
  \bibinfo{journal}{Phys. Rev. Lett.} \textbf{\bibinfo{volume}{111}},
  \bibinfo{pages}{234101} (\bibinfo{year}{2013}),
  \urlprefix\url{https://link.aps.org/doi/10.1103/PhysRevLett.111.234101}.

\bibitem[{\citenamefont{Lee et~al.}(2014)\citenamefont{Lee, Chan, and
  Wang}}]{Lee-2014}
\bibinfo{author}{\bibfnamefont{T.~E.} \bibnamefont{Lee}},
  \bibinfo{author}{\bibfnamefont{C.-K.} \bibnamefont{Chan}}, \bibnamefont{and}
  \bibinfo{author}{\bibfnamefont{S.}~\bibnamefont{Wang}},
  \bibinfo{journal}{Phys. Rev. E} \textbf{\bibinfo{volume}{89}},
  \bibinfo{pages}{022913} (\bibinfo{year}{2014}),
  \urlprefix\url{https://link.aps.org/doi/10.1103/PhysRevE.89.022913}.

\bibitem[{\citenamefont{Walter et~al.}(2014)\citenamefont{Walter, Nunnenkamp,
  and Bruder}}]{Walter-2014}
\bibinfo{author}{\bibfnamefont{S.}~\bibnamefont{Walter}},
  \bibinfo{author}{\bibfnamefont{A.}~\bibnamefont{Nunnenkamp}},
  \bibnamefont{and} \bibinfo{author}{\bibfnamefont{C.}~\bibnamefont{Bruder}},
  \bibinfo{journal}{Phys. Rev. Lett.} \textbf{\bibinfo{volume}{112}},
  \bibinfo{pages}{094102} (\bibinfo{year}{2014}),
  \urlprefix\url{https://link.aps.org/doi/10.1103/PhysRevLett.112.094102}.

\bibitem[{\citenamefont{Walter et~al.}(2015)\citenamefont{Walter, Nunnenkamp,
  and Bruder}}]{Walter-2015}
\bibinfo{author}{\bibfnamefont{S.}~\bibnamefont{Walter}},
  \bibinfo{author}{\bibfnamefont{A.}~\bibnamefont{Nunnenkamp}},
  \bibnamefont{and} \bibinfo{author}{\bibfnamefont{C.}~\bibnamefont{Bruder}},
  \bibinfo{journal}{Annalen der Physik} \textbf{\bibinfo{volume}{527}},
  \bibinfo{pages}{131} (\bibinfo{year}{2015}), ISSN \bibinfo{issn}{1521-3889},
  \urlprefix\url{http://dx.doi.org/10.1002/andp.201400144}.

\bibitem[{\citenamefont{Ameri et~al.}(2015)\citenamefont{Ameri, Eghbali-Arani,
  Mari, Farace, Kheirandish, Giovannetti, and Fazio}}]{Ameri-2015}
\bibinfo{author}{\bibfnamefont{V.}~\bibnamefont{Ameri}},
  \bibinfo{author}{\bibfnamefont{M.}~\bibnamefont{Eghbali-Arani}},
  \bibinfo{author}{\bibfnamefont{A.}~\bibnamefont{Mari}},
  \bibinfo{author}{\bibfnamefont{A.}~\bibnamefont{Farace}},
  \bibinfo{author}{\bibfnamefont{F.}~\bibnamefont{Kheirandish}},
  \bibinfo{author}{\bibfnamefont{V.}~\bibnamefont{Giovannetti}},
  \bibnamefont{and} \bibinfo{author}{\bibfnamefont{R.}~\bibnamefont{Fazio}},
  \bibinfo{journal}{Phys. Rev. A} \textbf{\bibinfo{volume}{91}},
  \bibinfo{pages}{012301} (\bibinfo{year}{2015}),
  \urlprefix\url{https://link.aps.org/doi/10.1103/PhysRevA.91.012301}.

\bibitem[{\citenamefont{Weiss et~al.}(2016)\citenamefont{Weiss, Kronwald, and
  Marquardt}}]{Weiss-2016}
\bibinfo{author}{\bibfnamefont{T.}~\bibnamefont{Weiss}},
  \bibinfo{author}{\bibfnamefont{A.}~\bibnamefont{Kronwald}}, \bibnamefont{and}
  \bibinfo{author}{\bibfnamefont{F.}~\bibnamefont{Marquardt}},
  \bibinfo{journal}{New Journal of Physics} \textbf{\bibinfo{volume}{18}},
  \bibinfo{pages}{013043} (\bibinfo{year}{2016}),
  \urlprefix\url{http://stacks.iop.org/1367-2630/18/i=1/a=013043}.

\bibitem[{\citenamefont{Fiderer et~al.}(2016)\citenamefont{Fiderer,
  Ku\ifmmode~\acute{s}\else \'{s}\fi{}, and Braun}}]{Fiderer-2016}
\bibinfo{author}{\bibfnamefont{L.~J.} \bibnamefont{Fiderer}},
  \bibinfo{author}{\bibfnamefont{M.}~\bibnamefont{Ku\ifmmode~\acute{s}\else
  \'{s}\fi{}}}, \bibnamefont{and}
  \bibinfo{author}{\bibfnamefont{D.}~\bibnamefont{Braun}},
  \bibinfo{journal}{Phys. Rev. A} \textbf{\bibinfo{volume}{94}},
  \bibinfo{pages}{032336} (\bibinfo{year}{2016}),
  \urlprefix\url{https://link.aps.org/doi/10.1103/PhysRevA.94.032336}.

\bibitem[{\citenamefont{L\"orch et~al.}(2016)\citenamefont{L\"orch, Amitai,
  Nunnenkamp, and Bruder}}]{Loerch-2016}
\bibinfo{author}{\bibfnamefont{N.}~\bibnamefont{L\"orch}},
  \bibinfo{author}{\bibfnamefont{E.}~\bibnamefont{Amitai}},
  \bibinfo{author}{\bibfnamefont{A.}~\bibnamefont{Nunnenkamp}},
  \bibnamefont{and} \bibinfo{author}{\bibfnamefont{C.}~\bibnamefont{Bruder}},
  \bibinfo{journal}{Phys. Rev. Lett.} \textbf{\bibinfo{volume}{117}},
  \bibinfo{pages}{073601} (\bibinfo{year}{2016}),
  \urlprefix\url{https://link.aps.org/doi/10.1103/PhysRevLett.117.073601}.

\bibitem[{\citenamefont{L\"orch et~al.}(2017)\citenamefont{L\"orch, Nigg,
  Nunnenkamp, Tiwari, and Bruder}}]{Loerch-2017}
\bibinfo{author}{\bibfnamefont{N.}~\bibnamefont{L\"orch}},
  \bibinfo{author}{\bibfnamefont{S.~E.} \bibnamefont{Nigg}},
  \bibinfo{author}{\bibfnamefont{A.}~\bibnamefont{Nunnenkamp}},
  \bibinfo{author}{\bibfnamefont{R.~P.} \bibnamefont{Tiwari}},
  \bibnamefont{and} \bibinfo{author}{\bibfnamefont{C.}~\bibnamefont{Bruder}},
  \bibinfo{journal}{Phys. Rev. Lett.} \textbf{\bibinfo{volume}{118}},
  \bibinfo{pages}{243602} (\bibinfo{year}{2017}),
  \urlprefix\url{https://link.aps.org/doi/10.1103/PhysRevLett.118.243602}.

\bibitem[{\citenamefont{Makino et~al.}(2016)\citenamefont{Makino, Hashimoto,
  Yoshikawa, Ohdan, Toyama, van Loock, and Furusawa}}]{Makino-2016}
\bibinfo{author}{\bibfnamefont{K.}~\bibnamefont{Makino}},
  \bibinfo{author}{\bibfnamefont{Y.}~\bibnamefont{Hashimoto}},
  \bibinfo{author}{\bibfnamefont{J.-i.} \bibnamefont{Yoshikawa}},
  \bibinfo{author}{\bibfnamefont{H.}~\bibnamefont{Ohdan}},
  \bibinfo{author}{\bibfnamefont{T.}~\bibnamefont{Toyama}},
  \bibinfo{author}{\bibfnamefont{P.}~\bibnamefont{van Loock}},
  \bibnamefont{and} \bibinfo{author}{\bibfnamefont{A.}~\bibnamefont{Furusawa}},
  \bibinfo{journal}{Science Advances} \textbf{\bibinfo{volume}{2}}
  (\bibinfo{year}{2016}),
  \eprint{http://advances.sciencemag.org/content/2/5/e1501772.full.pdf},
  \urlprefix\url{http://advances.sciencemag.org/content/2/5/e1501772}.

\bibitem[{\citenamefont{Giorgi et~al.}(2016)\citenamefont{Giorgi, Galve, and
  Zambrini}}]{Giorgi-2016}
\bibinfo{author}{\bibfnamefont{G.~L.} \bibnamefont{Giorgi}},
  \bibinfo{author}{\bibfnamefont{F.}~\bibnamefont{Galve}}, \bibnamefont{and}
  \bibinfo{author}{\bibfnamefont{R.}~\bibnamefont{Zambrini}},
  \bibinfo{journal}{Phys. Rev. A} \textbf{\bibinfo{volume}{94}},
  \bibinfo{pages}{052121} (\bibinfo{year}{2016}),
  \urlprefix\url{https://link.aps.org/doi/10.1103/PhysRevA.94.052121}.

\bibitem[{\citenamefont{Bellomo et~al.}(2017)\citenamefont{Bellomo, Giorgi,
  Palma, and Zambrini}}]{Bellomo-2017}
\bibinfo{author}{\bibfnamefont{B.}~\bibnamefont{Bellomo}},
  \bibinfo{author}{\bibfnamefont{G.~L.} \bibnamefont{Giorgi}},
  \bibinfo{author}{\bibfnamefont{G.~M.} \bibnamefont{Palma}}, \bibnamefont{and}
  \bibinfo{author}{\bibfnamefont{R.}~\bibnamefont{Zambrini}},
  \bibinfo{journal}{Phys. Rev. A} \textbf{\bibinfo{volume}{95}},
  \bibinfo{pages}{043807} (\bibinfo{year}{2017}),
  \urlprefix\url{https://link.aps.org/doi/10.1103/PhysRevA.95.043807}.

\bibitem[{\citenamefont{Kirchmair et~al.}(2013)\citenamefont{Kirchmair,
  Vlastakis, Leghtas, Nigg, Paik, Ginossar, Mirrahimi, Frunzio, Girvin, and
  Schoelkopf}}]{Kirchmair-2013a}
\bibinfo{author}{\bibfnamefont{G.}~\bibnamefont{Kirchmair}},
  \bibinfo{author}{\bibfnamefont{B.}~\bibnamefont{Vlastakis}},
  \bibinfo{author}{\bibfnamefont{Z.}~\bibnamefont{Leghtas}},
  \bibinfo{author}{\bibfnamefont{S.~E.} \bibnamefont{Nigg}},
  \bibinfo{author}{\bibfnamefont{H.}~\bibnamefont{Paik}},
  \bibinfo{author}{\bibfnamefont{E.}~\bibnamefont{Ginossar}},
  \bibinfo{author}{\bibfnamefont{M.}~\bibnamefont{Mirrahimi}},
  \bibinfo{author}{\bibfnamefont{L.}~\bibnamefont{Frunzio}},
  \bibinfo{author}{\bibfnamefont{S.~M.} \bibnamefont{Girvin}},
  \bibnamefont{and} \bibinfo{author}{\bibfnamefont{R.~J.}
  \bibnamefont{Schoelkopf}}, \bibinfo{journal}{Nature}
  \textbf{\bibinfo{volume}{495}}, \bibinfo{pages}{205} (\bibinfo{year}{2013}).

\bibitem[{\citenamefont{Devoret and Schoelkopf}(2013)}]{Devoret-2013}
\bibinfo{author}{\bibfnamefont{M.~H.} \bibnamefont{Devoret}} \bibnamefont{and}
  \bibinfo{author}{\bibfnamefont{R.~J.} \bibnamefont{Schoelkopf}},
  \bibinfo{journal}{Science} \textbf{\bibinfo{volume}{339}},
  \bibinfo{pages}{1169} (\bibinfo{year}{2013}), ISSN \bibinfo{issn}{0036-8075},
  \eprint{http://science.sciencemag.org/content/339/6124/1169.full.pdf},
  \urlprefix\url{http://science.sciencemag.org/content/339/6124/1169}.

\bibitem[{\citenamefont{Anderson et~al.}(2016)\citenamefont{Anderson, Ma,
  Owens, Schuster, and Simon}}]{Anderson-2016}
\bibinfo{author}{\bibfnamefont{B.~M.} \bibnamefont{Anderson}},
  \bibinfo{author}{\bibfnamefont{R.}~\bibnamefont{Ma}},
  \bibinfo{author}{\bibfnamefont{C.}~\bibnamefont{Owens}},
  \bibinfo{author}{\bibfnamefont{D.~I.} \bibnamefont{Schuster}},
  \bibnamefont{and} \bibinfo{author}{\bibfnamefont{J.}~\bibnamefont{Simon}},
  \bibinfo{journal}{Phys. Rev. X} \textbf{\bibinfo{volume}{6}},
  \bibinfo{pages}{041043} (\bibinfo{year}{2016}),
  \urlprefix\url{https://link.aps.org/doi/10.1103/PhysRevX.6.041043}.

\bibitem[{\citenamefont{Shankar et~al.}(2013)\citenamefont{Shankar, Hatridge,
  Leghtas, Sliwa, Narla, Vool, Girvin, Frunzio, Mirrahimi, and
  Devoret}}]{Shankar-2013}
\bibinfo{author}{\bibfnamefont{S.}~\bibnamefont{Shankar}},
  \bibinfo{author}{\bibfnamefont{M.}~\bibnamefont{Hatridge}},
  \bibinfo{author}{\bibfnamefont{Z.}~\bibnamefont{Leghtas}},
  \bibinfo{author}{\bibfnamefont{K.~M.} \bibnamefont{Sliwa}},
  \bibinfo{author}{\bibfnamefont{A.}~\bibnamefont{Narla}},
  \bibinfo{author}{\bibfnamefont{U.}~\bibnamefont{Vool}},
  \bibinfo{author}{\bibfnamefont{S.~M.} \bibnamefont{Girvin}},
  \bibinfo{author}{\bibfnamefont{L.}~\bibnamefont{Frunzio}},
  \bibinfo{author}{\bibfnamefont{M.}~\bibnamefont{Mirrahimi}},
  \bibnamefont{and} \bibinfo{author}{\bibfnamefont{M.~H.}
  \bibnamefont{Devoret}}, \bibinfo{journal}{Nature}
  \textbf{\bibinfo{volume}{504}}, \bibinfo{pages}{419} (\bibinfo{year}{2013}),
  ISSN \bibinfo{issn}{0028-0836}, \bibinfo{note}{letter},
  \urlprefix\url{http://dx.doi.org/10.1038/nature12802}.

\bibitem[{\citenamefont{Kimchi-Schwartz
  et~al.}(2016)\citenamefont{Kimchi-Schwartz, Martin, Flurin, Aron, Kulkarni,
  Tureci, and Siddiqi}}]{Kimchi-Schwartz-2016}
\bibinfo{author}{\bibfnamefont{M.~E.} \bibnamefont{Kimchi-Schwartz}},
  \bibinfo{author}{\bibfnamefont{L.}~\bibnamefont{Martin}},
  \bibinfo{author}{\bibfnamefont{E.}~\bibnamefont{Flurin}},
  \bibinfo{author}{\bibfnamefont{C.}~\bibnamefont{Aron}},
  \bibinfo{author}{\bibfnamefont{M.}~\bibnamefont{Kulkarni}},
  \bibinfo{author}{\bibfnamefont{H.~E.} \bibnamefont{Tureci}},
  \bibnamefont{and} \bibinfo{author}{\bibfnamefont{I.}~\bibnamefont{Siddiqi}},
  \bibinfo{journal}{Phys. Rev. Lett.} \textbf{\bibinfo{volume}{116}},
  \bibinfo{pages}{240503} (\bibinfo{year}{2016}),
  \urlprefix\url{https://link.aps.org/doi/10.1103/PhysRevLett.116.240503}.

\bibitem[{\citenamefont{Liu et~al.}(2016)\citenamefont{Liu, Shankar, Ofek,
  Hatridge, Narla, Sliwa, Frunzio, Schoelkopf, and Devoret}}]{Liu-2016}
\bibinfo{author}{\bibfnamefont{Y.}~\bibnamefont{Liu}},
  \bibinfo{author}{\bibfnamefont{S.}~\bibnamefont{Shankar}},
  \bibinfo{author}{\bibfnamefont{N.}~\bibnamefont{Ofek}},
  \bibinfo{author}{\bibfnamefont{M.}~\bibnamefont{Hatridge}},
  \bibinfo{author}{\bibfnamefont{A.}~\bibnamefont{Narla}},
  \bibinfo{author}{\bibfnamefont{K.~M.} \bibnamefont{Sliwa}},
  \bibinfo{author}{\bibfnamefont{L.}~\bibnamefont{Frunzio}},
  \bibinfo{author}{\bibfnamefont{R.~J.} \bibnamefont{Schoelkopf}},
  \bibnamefont{and} \bibinfo{author}{\bibfnamefont{M.~H.}
  \bibnamefont{Devoret}}, \bibinfo{journal}{Phys. Rev. X}
  \textbf{\bibinfo{volume}{6}}, \bibinfo{pages}{011022} (\bibinfo{year}{2016}),
  \urlprefix\url{https://link.aps.org/doi/10.1103/PhysRevX.6.011022}.

\bibitem[{\citenamefont{Albert et~al.}(2016)\citenamefont{Albert, Bradlyn,
  Fraas, and Jiang}}]{Albert-2016}
\bibinfo{author}{\bibfnamefont{V.~V.} \bibnamefont{Albert}},
  \bibinfo{author}{\bibfnamefont{B.}~\bibnamefont{Bradlyn}},
  \bibinfo{author}{\bibfnamefont{M.}~\bibnamefont{Fraas}}, \bibnamefont{and}
  \bibinfo{author}{\bibfnamefont{L.}~\bibnamefont{Jiang}},
  \bibinfo{journal}{Phys. Rev. X} \textbf{\bibinfo{volume}{6}},
  \bibinfo{pages}{041031} (\bibinfo{year}{2016}),
  \urlprefix\url{https://link.aps.org/doi/10.1103/PhysRevX.6.041031}.

\bibitem[{\citenamefont{Toyli et~al.}(2016)\citenamefont{Toyli, Eddins, Boutin,
  Puri, Hover, Bolkhovsky, Oliver, Blais, and Siddiqi}}]{Toyli-2016}
\bibinfo{author}{\bibfnamefont{D.~M.} \bibnamefont{Toyli}},
  \bibinfo{author}{\bibfnamefont{A.~W.} \bibnamefont{Eddins}},
  \bibinfo{author}{\bibfnamefont{S.}~\bibnamefont{Boutin}},
  \bibinfo{author}{\bibfnamefont{S.}~\bibnamefont{Puri}},
  \bibinfo{author}{\bibfnamefont{D.}~\bibnamefont{Hover}},
  \bibinfo{author}{\bibfnamefont{V.}~\bibnamefont{Bolkhovsky}},
  \bibinfo{author}{\bibfnamefont{W.~D.} \bibnamefont{Oliver}},
  \bibinfo{author}{\bibfnamefont{A.}~\bibnamefont{Blais}}, \bibnamefont{and}
  \bibinfo{author}{\bibfnamefont{I.}~\bibnamefont{Siddiqi}},
  \bibinfo{journal}{Phys. Rev. X} \textbf{\bibinfo{volume}{6}},
  \bibinfo{pages}{031004} (\bibinfo{year}{2016}),
  \urlprefix\url{https://link.aps.org/doi/10.1103/PhysRevX.6.031004}.

\bibitem[{\citenamefont{Pikovsky et~al.}(2003)\citenamefont{Pikovsky,
  Rosenblum, and Kurths}}]{Pikovsky}
\bibinfo{author}{\bibfnamefont{A.}~\bibnamefont{Pikovsky}},
  \bibinfo{author}{\bibfnamefont{M.}~\bibnamefont{Rosenblum}},
  \bibnamefont{and} \bibinfo{author}{\bibfnamefont{J.}~\bibnamefont{Kurths}},
  \emph{\bibinfo{title}{Synchronization: A Universal Concept in Nonlinear
  Sciences}} (\bibinfo{publisher}{Cambridge University Press},
  \bibinfo{year}{2003}).

\bibitem[{\citenamefont{Manzano et~al.}(2013)\citenamefont{Manzano, Galve,
  Giorgi, Hernandez-Garcia, and Zambrini}}]{Manzano-2013}
\bibinfo{author}{\bibfnamefont{G.}~\bibnamefont{Manzano}},
  \bibinfo{author}{\bibfnamefont{F.}~\bibnamefont{Galve}},
  \bibinfo{author}{\bibfnamefont{G.~L.} \bibnamefont{Giorgi}},
  \bibinfo{author}{\bibfnamefont{E.}~\bibnamefont{Hernandez-Garcia}},
  \bibnamefont{and} \bibinfo{author}{\bibfnamefont{R.}~\bibnamefont{Zambrini}},
  \bibinfo{journal}{Scientific Reports} \textbf{\bibinfo{volume}{3}},
  \bibinfo{pages}{1439 EP } (\bibinfo{year}{2013}), \bibinfo{note}{article},
  \urlprefix\url{http://dx.doi.org/10.1038/srep01439}.

\bibitem[{\citenamefont{Witthaut et~al.}(2017)\citenamefont{Witthaut,
  Wimberger, Burioni, and Timme}}]{Witthaut-2017}
\bibinfo{author}{\bibfnamefont{D.}~\bibnamefont{Witthaut}},
  \bibinfo{author}{\bibfnamefont{S.}~\bibnamefont{Wimberger}},
  \bibinfo{author}{\bibfnamefont{R.}~\bibnamefont{Burioni}}, \bibnamefont{and}
  \bibinfo{author}{\bibfnamefont{M.}~\bibnamefont{Timme}},
  \bibinfo{journal}{Nature Communications} \textbf{\bibinfo{volume}{8}},
  \bibinfo{pages}{14829 EP } (\bibinfo{year}{2017}), \bibinfo{note}{article},
  \urlprefix\url{http://dx.doi.org/10.1038/ncomms14829}.

\bibitem[{\citenamefont{Rips et~al.}(2012)\citenamefont{Rips, Kiffner,
  Wilson-Rae, and Hartmann}}]{Rips-2012}
\bibinfo{author}{\bibfnamefont{S.}~\bibnamefont{Rips}},
  \bibinfo{author}{\bibfnamefont{M.}~\bibnamefont{Kiffner}},
  \bibinfo{author}{\bibfnamefont{I.}~\bibnamefont{Wilson-Rae}},
  \bibnamefont{and} \bibinfo{author}{\bibfnamefont{M.~J.}
  \bibnamefont{Hartmann}}, \bibinfo{journal}{New Journal of Physics}
  \textbf{\bibinfo{volume}{14}}, \bibinfo{pages}{023042}
  (\bibinfo{year}{2012}),
  \urlprefix\url{http://stacks.iop.org/1367-2630/14/i=2/a=023042}.

\bibitem[{\citenamefont{Nigg et~al.}(2012)\citenamefont{Nigg, Paik, Vlastakis,
  Kirchmair, Shankar, Frunzio, Devoret, Schoelkopf, and Girvin}}]{Nigg-2012a}
\bibinfo{author}{\bibfnamefont{S.~E.} \bibnamefont{Nigg}},
  \bibinfo{author}{\bibfnamefont{H.}~\bibnamefont{Paik}},
  \bibinfo{author}{\bibfnamefont{B.}~\bibnamefont{Vlastakis}},
  \bibinfo{author}{\bibfnamefont{G.}~\bibnamefont{Kirchmair}},
  \bibinfo{author}{\bibfnamefont{S.}~\bibnamefont{Shankar}},
  \bibinfo{author}{\bibfnamefont{L.}~\bibnamefont{Frunzio}},
  \bibinfo{author}{\bibfnamefont{M.~H.} \bibnamefont{Devoret}},
  \bibinfo{author}{\bibfnamefont{R.~J.} \bibnamefont{Schoelkopf}},
  \bibnamefont{and} \bibinfo{author}{\bibfnamefont{S.~M.}
  \bibnamefont{Girvin}}, \bibinfo{journal}{Phys. Rev. Lett.}
  \textbf{\bibinfo{volume}{108}}, \bibinfo{pages}{240502}
  (\bibinfo{year}{2012}),
  \urlprefix\url{http://link.aps.org/doi/10.1103/PhysRevLett.108.240502}.

\bibitem[{\citenamefont{Bourassa et~al.}(2012)\citenamefont{Bourassa, Beaudoin,
  Gambetta, and Blais}}]{Bourassa-2012b}
\bibinfo{author}{\bibfnamefont{J.}~\bibnamefont{Bourassa}},
  \bibinfo{author}{\bibfnamefont{F.}~\bibnamefont{Beaudoin}},
  \bibinfo{author}{\bibfnamefont{J.~M.} \bibnamefont{Gambetta}},
  \bibnamefont{and} \bibinfo{author}{\bibfnamefont{A.}~\bibnamefont{Blais}},
  \bibinfo{journal}{Phys. Rev. A} \textbf{\bibinfo{volume}{86}},
  \bibinfo{pages}{013814} (\bibinfo{year}{2012}),
  \urlprefix\url{http://link.aps.org/doi/10.1103/PhysRevA.86.013814}.

\bibitem[{\citenamefont{Solgun et~al.}(2014)\citenamefont{Solgun, Abraham, and
  DiVincenzo}}]{Solgun-2014}
\bibinfo{author}{\bibfnamefont{F.}~\bibnamefont{Solgun}},
  \bibinfo{author}{\bibfnamefont{D.~W.} \bibnamefont{Abraham}},
  \bibnamefont{and} \bibinfo{author}{\bibfnamefont{D.~P.}
  \bibnamefont{DiVincenzo}}, \bibinfo{journal}{Phys. Rev. B}
  \textbf{\bibinfo{volume}{90}}, \bibinfo{pages}{134504}
  (\bibinfo{year}{2014}),
  \urlprefix\url{https://link.aps.org/doi/10.1103/PhysRevB.90.134504}.

\bibitem[{\citenamefont{Smith et~al.}(2016)\citenamefont{Smith, Kou, Vool, Pop,
  Frunzio, Schoelkopf, and Devoret}}]{Smith-2016}
\bibinfo{author}{\bibfnamefont{W.~C.} \bibnamefont{Smith}},
  \bibinfo{author}{\bibfnamefont{A.}~\bibnamefont{Kou}},
  \bibinfo{author}{\bibfnamefont{U.}~\bibnamefont{Vool}},
  \bibinfo{author}{\bibfnamefont{I.~M.} \bibnamefont{Pop}},
  \bibinfo{author}{\bibfnamefont{L.}~\bibnamefont{Frunzio}},
  \bibinfo{author}{\bibfnamefont{R.~J.} \bibnamefont{Schoelkopf}},
  \bibnamefont{and} \bibinfo{author}{\bibfnamefont{M.~H.}
  \bibnamefont{Devoret}}, \bibinfo{journal}{Phys. Rev. B}
  \textbf{\bibinfo{volume}{94}}, \bibinfo{pages}{144507}
  (\bibinfo{year}{2016}),
  \urlprefix\url{https://link.aps.org/doi/10.1103/PhysRevB.94.144507}.

\bibitem[{\citenamefont{Malekakhlagh et~al.}(2016)\citenamefont{Malekakhlagh,
  Petrescu, and T\"ureci}}]{Malekakhlagh-2016}
\bibinfo{author}{\bibfnamefont{M.}~\bibnamefont{Malekakhlagh}},
  \bibinfo{author}{\bibfnamefont{A.}~\bibnamefont{Petrescu}}, \bibnamefont{and}
  \bibinfo{author}{\bibfnamefont{H.~E.} \bibnamefont{T\"ureci}},
  \bibinfo{journal}{Phys. Rev. A} \textbf{\bibinfo{volume}{94}},
  \bibinfo{pages}{063848} (\bibinfo{year}{2016}),
  \urlprefix\url{https://link.aps.org/doi/10.1103/PhysRevA.94.063848}.

\bibitem[{sup()}]{supp_mat_qsync}
\bibinfo{note}{See ancillary files on the arxiv abstract page of the article.}

\bibitem[{\citenamefont{Holland et~al.}(2015)\citenamefont{Holland, Vlastakis,
  Heeres, Reagor, Vool, Leghtas, Frunzio, Kirchmair, Devoret, Mirrahimi
  et~al.}}]{Holland-2015}
\bibinfo{author}{\bibfnamefont{E.~T.} \bibnamefont{Holland}},
  \bibinfo{author}{\bibfnamefont{B.}~\bibnamefont{Vlastakis}},
  \bibinfo{author}{\bibfnamefont{R.~W.} \bibnamefont{Heeres}},
  \bibinfo{author}{\bibfnamefont{M.~J.} \bibnamefont{Reagor}},
  \bibinfo{author}{\bibfnamefont{U.}~\bibnamefont{Vool}},
  \bibinfo{author}{\bibfnamefont{Z.}~\bibnamefont{Leghtas}},
  \bibinfo{author}{\bibfnamefont{L.}~\bibnamefont{Frunzio}},
  \bibinfo{author}{\bibfnamefont{G.}~\bibnamefont{Kirchmair}},
  \bibinfo{author}{\bibfnamefont{M.~H.} \bibnamefont{Devoret}},
  \bibinfo{author}{\bibfnamefont{M.}~\bibnamefont{Mirrahimi}},
  \bibnamefont{et~al.}, \bibinfo{journal}{Phys. Rev. Lett.}
  \textbf{\bibinfo{volume}{115}}, \bibinfo{pages}{180501}
  (\bibinfo{year}{2015}),
  \urlprefix\url{https://link.aps.org/doi/10.1103/PhysRevLett.115.180501}.

\bibitem[{\citenamefont{Hush et~al.}(2015)\citenamefont{Hush, Li, Genway,
  Lesanovsky, and Armour}}]{Hush-2015}
\bibinfo{author}{\bibfnamefont{M.~R.} \bibnamefont{Hush}},
  \bibinfo{author}{\bibfnamefont{W.}~\bibnamefont{Li}},
  \bibinfo{author}{\bibfnamefont{S.}~\bibnamefont{Genway}},
  \bibinfo{author}{\bibfnamefont{I.}~\bibnamefont{Lesanovsky}},
  \bibnamefont{and} \bibinfo{author}{\bibfnamefont{A.~D.}
  \bibnamefont{Armour}}, \bibinfo{journal}{Phys. Rev. A}
  \textbf{\bibinfo{volume}{91}}, \bibinfo{pages}{061401}
  (\bibinfo{year}{2015}),
  \urlprefix\url{https://link.aps.org/doi/10.1103/PhysRevA.91.061401}.

\bibitem[{\citenamefont{Plenio}(2005)}]{Plenio-2005}
\bibinfo{author}{\bibfnamefont{M.~B.} \bibnamefont{Plenio}},
  \bibinfo{journal}{Phys. Rev. Lett.} \textbf{\bibinfo{volume}{95}},
  \bibinfo{pages}{090503} (\bibinfo{year}{2005}),
  \urlprefix\url{https://link.aps.org/doi/10.1103/PhysRevLett.95.090503}.

\bibitem[{\citenamefont{Wiseman and Milburn}(2009)}]{Wiseman_Milburn-2009}
\bibinfo{author}{\bibfnamefont{H.~M.} \bibnamefont{Wiseman}} \bibnamefont{and}
  \bibinfo{author}{\bibfnamefont{G.~J.} \bibnamefont{Milburn}},
  \emph{\bibinfo{title}{Quantum Measurement and Control}}
  (\bibinfo{publisher}{Cambridge Univ. Press}, \bibinfo{year}{2009}).

\bibitem[{\citenamefont{Benedetti et~al.}(2016)\citenamefont{Benedetti, Galve,
  Mandarino, Paris, and Zambrini}}]{Benedetti-2016}
\bibinfo{author}{\bibfnamefont{C.}~\bibnamefont{Benedetti}},
  \bibinfo{author}{\bibfnamefont{F.}~\bibnamefont{Galve}},
  \bibinfo{author}{\bibfnamefont{A.}~\bibnamefont{Mandarino}},
  \bibinfo{author}{\bibfnamefont{M.~G.~A.} \bibnamefont{Paris}},
  \bibnamefont{and} \bibinfo{author}{\bibfnamefont{R.}~\bibnamefont{Zambrini}},
  \bibinfo{journal}{Phys. Rev. A} \textbf{\bibinfo{volume}{94}},
  \bibinfo{pages}{052118} (\bibinfo{year}{2016}),
  \urlprefix\url{https://link.aps.org/doi/10.1103/PhysRevA.94.052118}.

\bibitem[{\citenamefont{Li et~al.}(2017)\citenamefont{Li, Li, and
  Song}}]{Li-2017}
\bibinfo{author}{\bibfnamefont{W.}~\bibnamefont{Li}},
  \bibinfo{author}{\bibfnamefont{C.}~\bibnamefont{Li}}, \bibnamefont{and}
  \bibinfo{author}{\bibfnamefont{H.}~\bibnamefont{Song}},
  \bibinfo{journal}{Phys. Rev. E} \textbf{\bibinfo{volume}{95}},
  \bibinfo{pages}{022204} (\bibinfo{year}{2017}),
  \urlprefix\url{https://link.aps.org/doi/10.1103/PhysRevE.95.022204}.

\bibitem[{\citenamefont{Nigg et~al.}(2017)\citenamefont{Nigg, L{\"o}rch, and
  Tiwari}}]{Nigg-2017}
\bibinfo{author}{\bibfnamefont{S.~E.} \bibnamefont{Nigg}},
  \bibinfo{author}{\bibfnamefont{N.}~\bibnamefont{L{\"o}rch}},
  \bibnamefont{and} \bibinfo{author}{\bibfnamefont{R.~P.}
  \bibnamefont{Tiwari}}, \bibinfo{journal}{Science Advances}
  \textbf{\bibinfo{volume}{3}} (\bibinfo{year}{2017}),
  \eprint{http://advances.sciencemag.org/content/3/4/e1602273.full.pdf},
  \urlprefix\url{http://advances.sciencemag.org/content/3/4/e1602273}.

\bibitem[{\citenamefont{Puri et~al.}(2017)\citenamefont{Puri, Andersen,
  Grimsmo, and Blais}}]{Puri-2017}
\bibinfo{author}{\bibfnamefont{S.}~\bibnamefont{Puri}},
  \bibinfo{author}{\bibfnamefont{C.~K.} \bibnamefont{Andersen}},
  \bibinfo{author}{\bibfnamefont{A.~L.} \bibnamefont{Grimsmo}},
  \bibnamefont{and} \bibinfo{author}{\bibfnamefont{A.}~\bibnamefont{Blais}},
  \textbf{\bibinfo{volume}{8}}, \bibinfo{pages}{15785 EP }
  (\bibinfo{year}{2017}), \bibinfo{note}{article},
  \urlprefix\url{http://dx.doi.org/10.1038/ncomms15785}.

\end{thebibliography}

\end{document}